\documentclass[useAMS,usenatbib]{mn2e}

\usepackage{float} 
\usepackage{epsfig}
\usepackage{graphicx}
\usepackage[latin1]{inputenc}
\usepackage{latexsym}
\def\simlt{\ \raise -2.truept\hbox{\rlap{\hbox{$\sim$}}\raise5.truept   %
\hbox{$<$}\ }}
\def\simgt{\ \raise -2.truept\hbox{\rlap{\hbox{$\sim$}}\raise5.truept   %
\hbox{$>$}\ }}                                                          %

\def\be{\begin{equation}}
\def\ee{\end{equation}}
\def\newline{\hfil\break}

\def\la{\mathrel{\hbox{\rlap{\hbox{\lower4pt\hbox{$\sim$}}}\hbox{$<$}}}}
\def\ga{\mathrel{\hbox{\rlap{\hbox{\lower4pt\hbox{$\sim$}}}\hbox{$>$}}}}

\def\MS7{MS 0735.6+7421}

\title[Non-thermal SZE in galaxy clusters cavities]{Thermal and non-thermal Sunyaev-Zel'dovich effect in the cavities of the galaxy cluster MS 0735.6+7421: the role of the thermal density in the cavity}
\author[P. Marchegiani]{P. Marchegiani$^{1,2}$\thanks{E-mail: paolo.marchegiani@uniroma1.it ; Paolo.Marchegiani@wits.ac.za}\\
$^{1}$Dipartimento di Fisica, Universit\`a Sapienza, P.le Aldo Moro 2, 00185, Roma, Italy\\
$^{2}$School of Physics, University of the Witwatersrand, Private Bag 3, 2050-Johannesburg, South Africa\\
}

\begin{document}

\date{Accepted 2021 March 06. Received 2021 February 08; in original form 2020 October 14}

\pagerange{\pageref{firstpage}--\pageref{lastpage}} \pubyear{2021}

\maketitle

\label{firstpage}

\begin{abstract}
The galaxy cluster MS 0735.6+7421 hosts two large X-ray cavities, filled with radio emission, where a decrease of the Sunyaev-Zel'dovich (SZ) effect has been detected, without establishing if its origin is thermal (from a gas with very high temperature) or non-thermal. In this paper we study how thermal and non-thermal contributions to the SZ effect in the cavities are related; in fact, Coulomb interactions with the thermal gas modify the spectrum of low energy non-thermal electrons, which dominate the non-thermal SZ effect; as a consequence, the intensity of the non-thermal SZ effect is stronger for lower density of the thermal gas inside the cavity. We calculate the non-thermal SZ effect in the cavities as a function of the thermal density, and compare the SZ effects produced by thermal and non-thermal components, and with the one from the external Intra Cluster Medium (ICM), searching for the best frequency range where it is possible to disentangle the different contributions. We find that for temperatures inside the cavities higher than $\sim1500$ keV the non-thermal SZ effect is expected to dominate on the thermal one, particularly at high frequencies ($\nu>500$ GHz), where it can also be a non-negligible fraction of the SZ effect from the external ICM. We also discuss the possible sources of astrophysical bias (as kinetic SZ effect and foreground emission from Galactic dust) and possible ways to address them, as well as necessary improvements in the modeling of the properties of cavities and the ICM.
\end{abstract}



\section{Introduction}     

X-ray cavities in galaxy clusters are deemed to be produced due to the expansion in the Intra Cluster Medium (ICM) of lobes inflated by relativistic jets produced by powerful Active Galactic Nuclei (AGN), usually located in the center of relaxed, cool-core clusters (e.g. Fabian 2012; Gitti, Brighenti \& McNamara 2012). 

When observed in the radio band, many cavities appear to be filled with a radio emission, while in other cases no radio emission is observed inside the cavities (see e.g. Birzan et al. 2020 for a recent observation of a sample of 42 systems of galaxy groups and clusters containing cavities, where it has been found that half of the objects do not host a radio emission). This fact is interpreted as due to a different stage of the evolutionary status of the non-thermal electrons in the lobes, with radio quiet cavities being observed at a late state of their evolution, when high-energy electrons producing the radio emission by synchrotron have already lost their energy because of radiative losses and adiabatic expansion.

It has been found that the energy stored in non-thermal electrons in cavities, as derived under the minimum energy assumption, is well below the energy required to inflate the cavity, by a factor that can be of the order of tens or a few hundreds (Ito et al. 2008). This finding suggests that a different population of particles, as non-thermal protons or a thermal gas, is storing most of the energy in the cavity. In the case of a thermal gas inside the cavity, the strong decrease of the X-ray emission indicates that this gas should have lower density compared to the surrounding ICM, and high temperature, of the order of hundreds of keV; the presence of such a gas is also predicted by relativistic hydrodynamic simulations studying the effect of the shock fronts produced by the interaction between the jets and the ICM (Prokhorov et al. 2012).  

A possible way to quantify the contribution to the total energy content of the cavity from different populations of electrons is provided by the Sunyaev-Zel'dovich (SZ) effect, i.e. the distortion on the Cosmic Microwave Background (CMB) spectrum produced by inverse Compton scattering in ionized media (Zel'dovich \& Sunyaev 1969). Since the SZ effects produced by thermal and non-thermal electrons have different spectral shapes (En\ss lin \& Kaiser 2000; Colafrancesco, Marchegiani \& Palladino 2003), it has been suggested that the SZ effect inside the cavity can be used to determine if the dominant population is the non-thermal or the high-temperature thermal one (Pfrommer, En\ss lin \& Sarazin 2005; Colafrancesco 2005; Prokhorov, Antonuccio-Delogu \& Silk 2010).

The first detection of the SZ effect in a cluster cavity has been reported by Abdulla et al. (2019), which observed the cluster \MS7 at 30 GHz with the CARMA interferometer, finding two cavities in the SZ map spatially coincident with the X-ray ones. From their analysis, they found that an SZ contribution coming from the plasma inside the cavities has been possibly detected, but, since the measure was performed only at one frequency, they have not been able to discriminate between a non-thermal electrons population with a low value of the normalized minimum momentum ($p\sim 1-10$), and a thermal population with temperature of the order of several hundreds or a few thousands keV (see their figure 8).

In this paper we study the relationship between the properties of the high-temperature thermal gas and the spectrum of the non-thermal electrons inside the cavities, and discuss possible ways to determine which of them provides the main contribution to the energy of the cavity and to the SZ effect. We use the fact that since non-thermal electrons in the cavities usually have steep spectra (Birzan et al. 2008), they are dominated in number by electrons located in the low energy part of the spectrum, which therefore produce most of the SZ effect (e.g. Colafrancesco \& Marchegiani 2011). In that spectral region the main contribution to the electrons energy losses is provided by Coulomb interactions with the electrons of the thermal gas (e.g. Sarazin 1999). Coulomb losses are basically not dependent on the electron energy (excluding the lowest energy region because of the $1/\beta$ dependence of the energy loss term), and linearly dependent on the gas density; since a high-temperature gas should have a low density in order to not be over-pressurized compared to the surrounding ICM, we can expect that in such a gas Coulomb losses should be reduced, and that as a consequence low energy non-thermal electrons should have a longer lifetime compared to the case of the denser ICM, resulting in a spectrum that can have a power-law shape until low values of the electrons energy, maximizing the non-thermal SZ effect.

Using the cluster \MS7 as a case of study, we follow the time evolution of the non-thermal electrons spectrum due to energy losses, adiabatic expansion and Fermi-II (re-)acceleration, with the aim to reproduce the shape of the observed radio spectrum produced by high energy electrons at a time of the order of the estimated age of the bubble, and we calculate the shape of the electrons spectrum at low energies, assuming different values for the density of the thermal gas inside the cavity. We first assume reference values for the thermal density, and later, assuming different temperature values for the thermal gas inside the cavity, we use the maximum values of the thermal density that do not produce a pressure in excess with respect to the external ICM. Using the resulting non-thermal electrons spectra we calculate the non-thermal SZ effect, and compare it with the thermal one produced by the high-temperature gas. Finally we compare the derived SZ effects with the one produced by the surrounding ICM, and discuss if it is possible to derive the properties of the cavities content from SZ measures, also at the light of astrophysical contaminations and approximations made in the calculations.

We assume a flat $\Lambda$CDM cosmology with $\Omega_m = 0.3$, $\Omega_{\Lambda} = 0.7$ and $H_0 =70$ km s$^{-1}$ Mpc$^{-1}$; for the redshift of the cluster \MS7, $z=0.216$, this cosmology model provides a luminosity distance of 1070 Mpc and a scale of 3.5 kpc/arcsec.

\section{Time evolution of the electrons spectrum}

\MS7 is a cool core cluster, where X-ray observations show the presence of two giant cavities (diameter $\sim200$ kpc) located at $\sim150$ kpc from the cluster center in opposite directions (McNamara et al. 2005; Gitti et al. 2007; Vantyghem et al. 2014). The cavities are filled with a radio emission, detected at 327 and 1400 MHz (Birzan et al. 2008) and recently at 143 MHz with LOFAR (Birzan et al. 2020). The observed high-frequency steepening of the radio spectrum suggests that electrons emitted in the jets from the central AGN are expanding inside the lobes and are losing their energy because of radiative losses and adiabatic expansion.

We follow the time evolution of the non-thermal electrons spectrum assuming that they are injected at an initial time with a power law spectrum in term of electrons normalized momentum $p=\beta\gamma$,
\begin{equation}
 N_e(p,t_0)=k_0 p^{-s}, 
\end{equation}
with $s=2.7$, and evolve with time subject to energy losses and possible Fermi-II (re-)acceleration produced by inhomogeneous magnetic fields, and without a source term after the injection, according to the equation
\begin{eqnarray}
\frac{\partial N_e(p)}{\partial t} & = & \frac{\partial}{\partial p} \left[ \left(-\frac{2}{p}D_{pp}+\sum_i b_i (p) \right) N_e(p) \right. \nonumber\\
 & & \left. + D_{pp}\frac{\partial N_e(p)}{\partial p}\right]
\label{evol.spettro}
\end{eqnarray}
(e.g. Schlickeiser 2002), where $D_{pp}$ is the diffusion coefficient in the momentum space, associated to Fermi-II acceleration processes, and the energy loss term $b(p)\equiv -dp/dt$ is given by the sum of Coulomb, bremsstrahlung, inverse Compton scattering and synchrotron losses (see e.g. Winner et al. 2019 for full expressions of these terms) and adiabatic losses. Energy losses are dominated at high energy by the radiative losses, proportional to $p^2$ and to the energy density of the CMB (for inverse Compton losses) and of the magnetic field (for synchrotron losses), while they are dominated at low energy by Coulomb losses, that are approximately constant with $p$, with the exception of the lowest energy part of the spectrum, where the presence of the $1/\beta$ factor produces a quick increase of the loss rate (e.g. Gould 1972), and are linearly dependent on the thermal gas density $n_{th}$.

The expansion of the bubble gives origin to adiabatic losses, described by an adjunctive loss term of the form 
\begin{equation}
b(p)=\frac{1}{3}\frac{1}{V}\frac{dV}{dt}p,
\end{equation}
where $V$ is the volume of the bubble. This results in a loss term dependent on the time, not only because of the change in the volume, but also because of the consequent changes of the magnetic field intensity and of the thermal density, with the magnetic field energy scaling as $u_B\propto V^{-4/3}$  (e.g. En\ss lin \& Gopal-Krishna 2001) and the thermal density scaling as $n_{th}\propto V^{-1}$. In the following we describe the volume changes as
\begin{equation}
V(t)=V_0 \left(\frac{t}{t_0}\right)^q ,
\end{equation}
and, with this assumption, the magnetic field and the thermal density also change as
\begin{eqnarray}
B(t) & = & B_0 \left(\frac{t}{t_0}\right)^{-\frac{2}{3}q} \\
n_{th} (t) & = & n_{th,0} \left(\frac{t}{t_0}\right)^{-q} .
\end{eqnarray}
We assume that the expansion starts at an initial time $t_0$ and continues until the time $\bar t$ at which the cluster is observed, fixing the value $ \bar t-t_0$ at the  age of the bubbles, which has been estimated to be of the order of 160 Myr using X-ray observations (Vantyghem et al. 2014) and MHD simulations (Ehlert et al. 2019). We model the expansion by using a Sedov-like exponent $q=6/5$ constant with time; we note that this is an approximation, because a more realistic modeling of the expansion should be described by an initial supersonic expansion phase, followed by a Sedov-like expansion, and by a slowdown when the internal bubble pressure becomes similar to the one of the external medium (see, e.g., discussion in En\ss lin \& Gopal-Krishna 2001). 

We assume that the magnetic field is constant inside the cavity normalizing its present value to the value estimated from the equipartition assumption, $B(\bar t)=4.7$ $\mu$G (Birzan et al. 2008), and we assume values of the thermal gas density at time $\bar t$ in the range $10^{-6}-10^{-3}$ cm$^{-3}$. We assume that non-thermal electrons reside inside a bubble with radius of 100 kpc, and have a constant density.

The last unknown parameter in eq.(\ref{evol.spettro}) is the diffusion coefficient in momentum space $D_{pp}$, which we write as $D_{pp}=\chi p^2 /4$, so that the associated acceleration characteristic time $\tau_{acc}=p^2/4D_{pp}$ (Brunetti \& Lazarian 2007) is equal to $\chi^{-1}$.

Eq.(\ref{evol.spettro}) is solved numerically adopting a finite difference scheme (Chang \& Cooper 1970; Park \& Petrosian 1996; Donnert \& Brunetti 2014) to obtain the evolution of the electrons spectrum as a function of the time. Once the electrons spectrum is calculated, we calculate the synchrotron spectrum produced at the time $\bar t$, leaving the normalization of the electron spectrum $k_0$ as a free parameter, and searching for the value of the $\chi$ parameter for which the resulting high frequency spectral steepening reproduces the observed one, using the data at 327 and 1400 MHz from Birzan et al. (2008), and the data at 143 MHz from Birzan et al. (2020). We use also the flux at 8500 MHz that in Birzan et al. (2008) is measured for the total galaxy emission, and not only in the lobes, considering it as an upper limit for the flux of lobes at that frequency.

\begin{figure}
\centering
\begin{tabular}{c}
\includegraphics[width=\columnwidth]{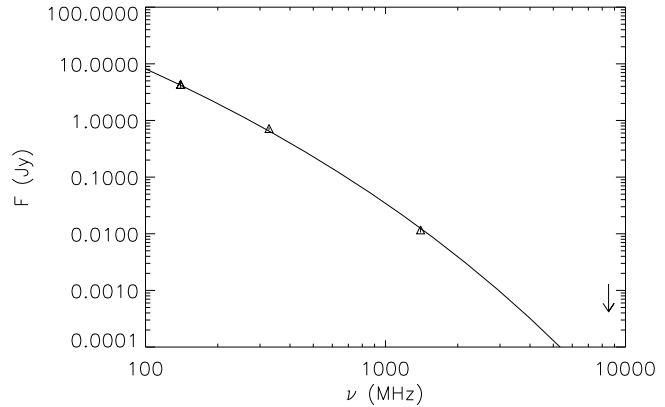}
\end{tabular}
\caption{Spectrum of the radio emission produced after 160 Myr from the injection. Data are from Birzan et al. (2008) and Birzan et al. (2020).}
\label{fig.radio}
\end{figure}

We find that the spectrum of the radio emission is well fitted after a time of $160$ Myr from the injection for a normalization of the electrons spectrum of $k_0=1.5\times10^{-3}$ cm$^{-3}$ and a reacceleration parameter of $\chi=3.5\times10^{-16}$ s$^{-1}$ (see Fig.\ref{fig.radio}). Since the fluxes used in this analysis are produced in both the lobes, in the following we approximate the non-thermal electrons density in a single bubble by dividing the value of $k_0$ by two.

\begin{figure}
\centering
\begin{tabular}{c}
\includegraphics[width=\columnwidth]{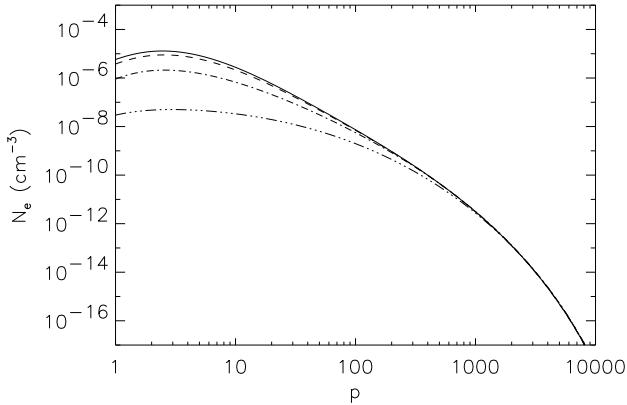}
\end{tabular}
\caption{Electrons spectrum after 160 Myr from the injection as a function of the electrons momentum for present day values of magnetic field of 4.7 $\mu$G and thermal density of $10^{-6}$ (solid line), $10^{-5}$ (dashed line), $10^{-4}$ (dot-dashed line), and $10^{-3}$ (three dots-dashed line) cm$^{-3}$.}
\label{fig.elettroni}
\end{figure}

While the electrons spectrum at high energies does not depend on the value of the density of the thermal gas inside the bubble, the shape of the electrons spectrum at low energies is instead strongly dependent on it because of the Coulomb losses. In Fig.\ref{fig.elettroni} we show 
the resulting spectra 
of the electrons after 160 Myr from the injection, with the normalization found from the fit to radio data, for four values of the thermal density at time $\bar t$. The electrons spectra overlap exactly at high energies, while they have different spectral shapes at low energies, where they also show a decrease for 
$p\simgt1$ because of the $1/\beta$ dependence of the Coulomb loss term. These different shapes do not impact on the synchrotron emission in the observed radio band, but have important consequences on the total number of non-thermal electrons, on their pressure and energy content, and on the SZ effect they produce, which is studied in next section.

\section{Non-thermal Sunyaev-Zel'dovich effect}

We calculate the SZ effect produced by the non-thermal electrons using the full relativistic formalism (Wright 1979; En\ss lin \& Kaiser 2000; Colafrancesco et al. 2003), where the intensity of the SZ effect at the normalized frequency $x=h\nu/(k_B T_{0})$, where $T_0$ is the CMB temperature, is written as:
\begin{equation}
\Delta I (x) =  \tau [J_1(x)-I_0(x)]
\end{equation}
where $I_0(x)$ is the unperturbed CMB spectrum,
\begin{equation}
I_0(x)=2\frac{ (k_B T_{0})^3}{(hc)^2}\frac{x^3}{e^x-1},
\end{equation}
$\tau$ is the optical depth, proportional to the integral along of the line of sight $\ell$ of the electrons density
\begin{equation}
\tau=\sigma_T \int n_e d\ell,
\label{optical.depth}
\end{equation}
with $\sigma_T$ being the Thomson cross section, and the function $J_1(x)$ is given by 
\begin{equation}
J_1(x)= \int_{-\infty}^{+\infty} I_0(xe^{-s}) P_1(s) ds,
\end{equation}
where $P_1(s)$ is the single scattering redistribution function, obtained by the integration
\begin{equation}
P_1(s)=\int_0^\infty f_e(p) P_s(s,p) dp,
\end{equation}
where $f_e(p)$ is the electrons spectrum 
normalized as $\int f_e(p) dp=1$, and $P_s(s,p)$ is the frequency redistribution function for a single electron with momentum $p$ (see En\ss lin \& Kaiser 2000 for an explicit form). 

With the described formalism it is possible to calculate the SZ in a full relativistic way for both thermal and non-thermal populations of electrons, using the appropriate shape of the spectrum $f_e(p)$. Usually, for non-thermal electrons this spectrum is written as a single or double power-law distribution with a minimum momentum $p_1$ that determines most of the SZ spectral properties (e.g. Colafrancesco et al. 2003), but it is possible to use any spectral shape, as the ones derived in previous section and shown in Fig.\ref{fig.elettroni},  
normalizing the spectrum to one in order to obtain the function $f_e(p)$. The electrons density $n_e$ at a given position is obtained integrating the spectra shown in Fig.\ref{fig.elettroni} over $p$, and the optical depth is derived from Eq.(\ref{optical.depth}), which is straightforward in our approach where the distribution of the electrons is assumed constant inside the bubble. 

\begin{figure}
\centering
\begin{tabular}{c}
\includegraphics[width=\columnwidth]{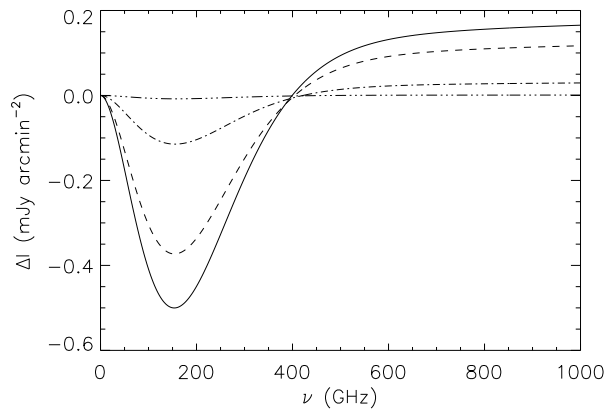}
\end{tabular}
\caption{Spectrum of the non-thermal SZ effect produced by electrons as in Fig.\ref{fig.elettroni} for different values of the thermal density:  $10^{-6}$ (solid line), $10^{-5}$ (dashed line), $10^{-4}$ (dot-dashed line), and $10^{-3}$ (three dots-dashed line) cm$^{-3}$.}
\label{fig.sz}
\end{figure}

For four values of the present day thermal density $n_{th}$ between $10^{-6}$ and $10^{-3}$ cm$^{-3}$ we calculate the non-thermal SZ effect, as well as the value of the pressure of the non-thermal electrons
\begin{equation}
P_{nt}=n_e \int_0^\infty f_e(p) \frac{1}{3} p \beta m_e c^2 dp
\end{equation}
and the corresponding pseudo-temperature 
\begin{equation}
\langle k_B T_e \rangle_{nt}= \int_0^\infty f_e(p) \frac{1}{3} p \beta m_e c^2 dp
\end{equation}
(En\ss lin \& Kaiser 2000), useful to estimate how much relativistic effects are important. The values of the non-thermal optical depth, pressure, and pseudo-temperature obtained in this way are reported in Table \ref{tab.sz}, while the resulting SZ spectra are shown in Fig.\ref{fig.sz}.

\begin{table}{}
\caption{Values of the thermal density assumed in the calculations, and corresponding values of non-thermal optical depth, pressure, and pseudo-temperature.}
\begin{center}
\begin{tabular}{|*{4}{c|}}
\hline 
$n_{th}$ & $\tau_{nt}$ & $P_{nt}$ & $\langle k_B T_e \rangle_{nt}$  \\
 cm$^{-3}$&  &  keV cm$^{-3}$ & keV  \\
\hline 
$10^{-6}$ & $1.81\times10^{-5}$ & $6.60\times10^{-2}$ & 1460 \\
$10^{-5}$ & $1.34\times10^{-5}$ & $5.39\times10^{-2}$ & 1613 \\
$10^{-4}$ & $4.01\times10^{-6}$ & $2.46\times10^{-2}$ & 2462 \\
$10^{-3}$ & $2.63\times10^{-7}$ & $5.05\times10^{-3}$ & 7710 \\
\hline
 \end{tabular}
 \end{center} 
 \label{tab.sz}
 \end{table}

It is possible to see that for low values of the thermal density the Coulomb losses are reduced, and the electrons spectrum remains similar to a power-law until low energies, approaching the case of a low value of the minimum momentum $p_1$ in the usual approach describing the electrons spectrum as a power-law with a low-energy cutoff. Since the density of non-thermal electrons is dominated by low-energy electrons, this results in higher values of optical depth and pressure, and as a consequence in a more intense SZ effect. Also, the dominance of low-energy electrons implies a lower value of the pseudo-temperature, which results in different spectral shapes; in fact, relativistic effects are more evident in the case of higher pseudo-temperature, as it can be seen by looking at the position of the crossover frequency, which is located at higher frequencies in the case of higher thermal density. Therefore, in principle it is possible to derive the properties of the thermal gas inside the bubble from the spectral shape of the non-thermal SZ effect.

\section{Comparison between non-thermal and thermal SZ effect}

In this section we compare the thermal and the non-thermal SZ effect inside the bubble, in order to establish which of them is expected to dominate. We assume several values of the temperature $T_e$ of the thermal gas inside the bubble, between 500 and 2000 keV, and assume a density of this gas such that its pressure does not exceed the pressure of the external ICM at the location of the cavity. This last quantity has been estimated to be $P_{cav}=6\times10^{-11}$ erg cm$^{-3}$ (Gitti et al. 2007), i.e. $3.75\times10^{-2}$ keV cm$^{-3}$. The density of the thermal gas inside the cavity is therefore assumed to be $n_{th}=P_{cav}/(k_BT_e)$. Using these values of the thermal gas density, we calculate the non-thermal electrons spectrum after 160 Myr from the injection as described in the previous section.

\begin{figure*}
\centering
\begin{tabular}{cc}
\includegraphics[width=\columnwidth]{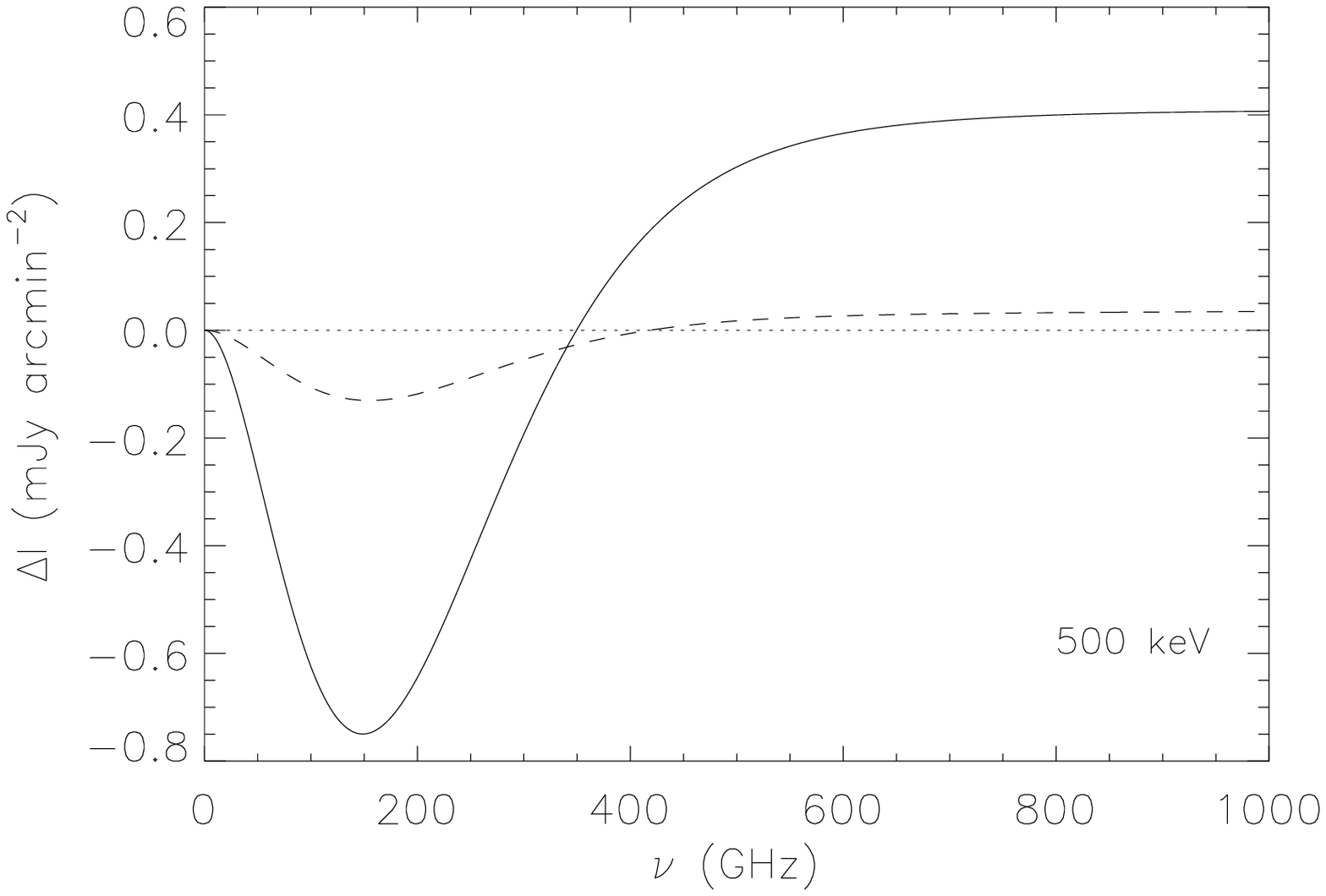} &
\includegraphics[width=\columnwidth]{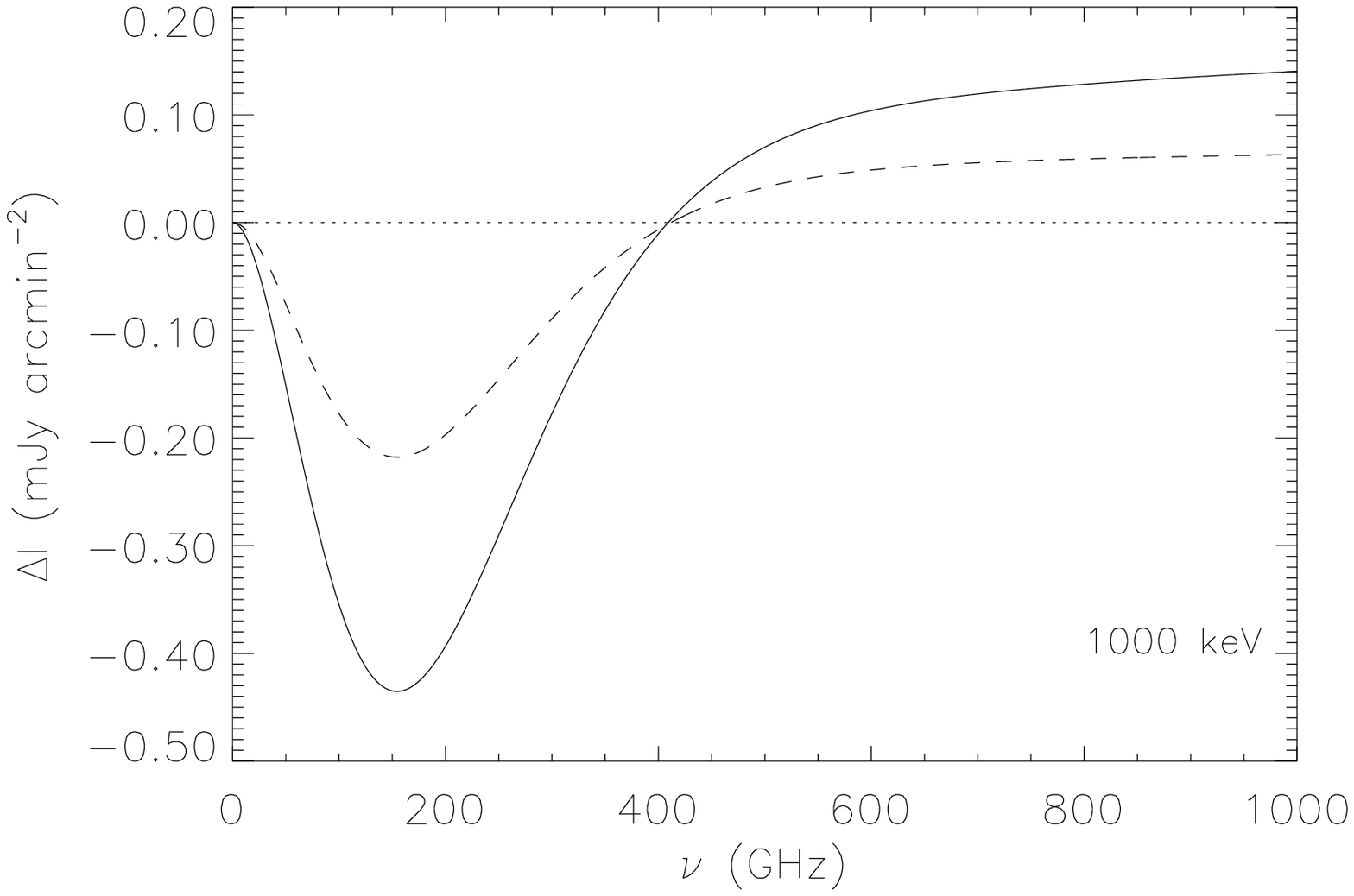} \\
\includegraphics[width=\columnwidth]{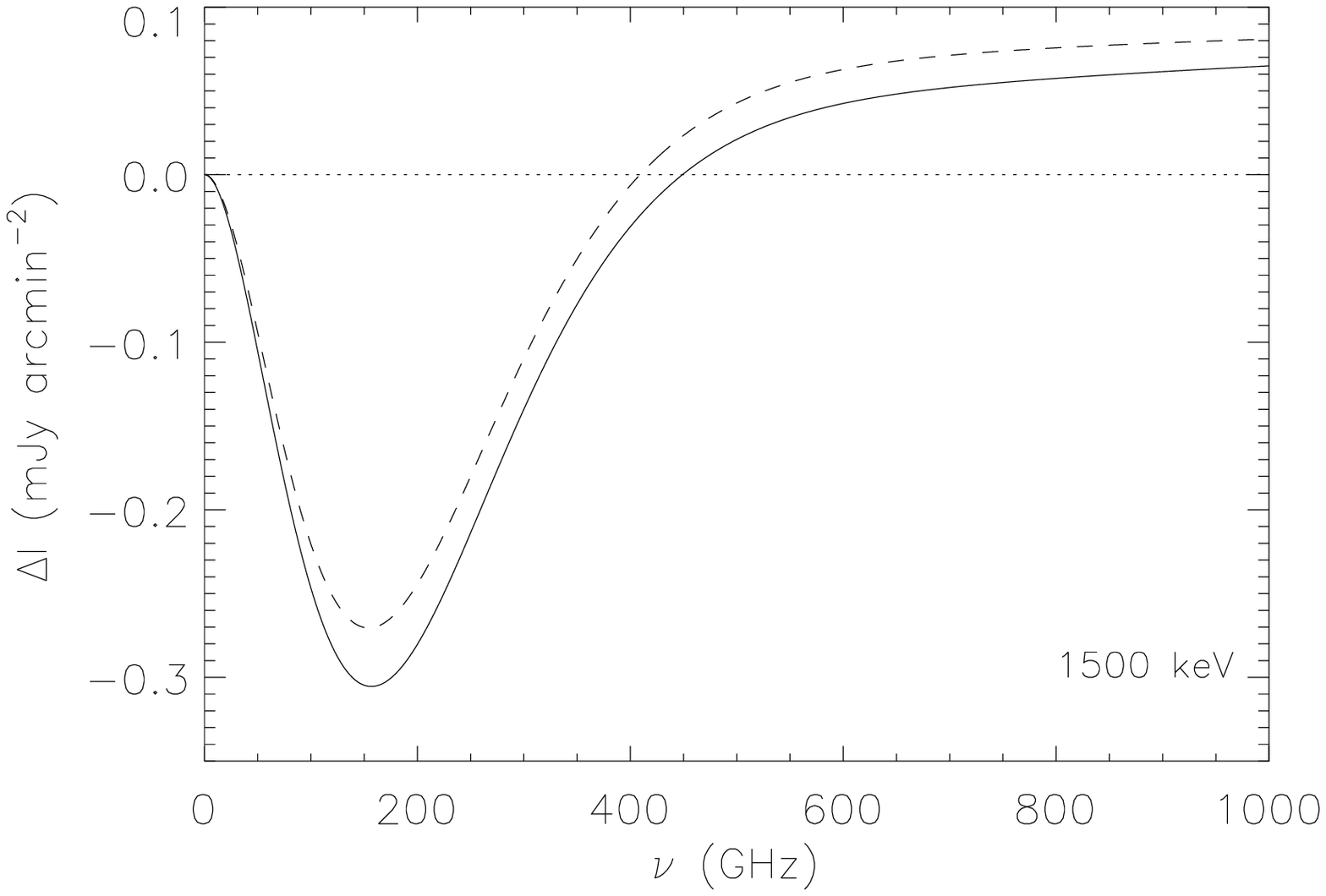} &
\includegraphics[width=\columnwidth]{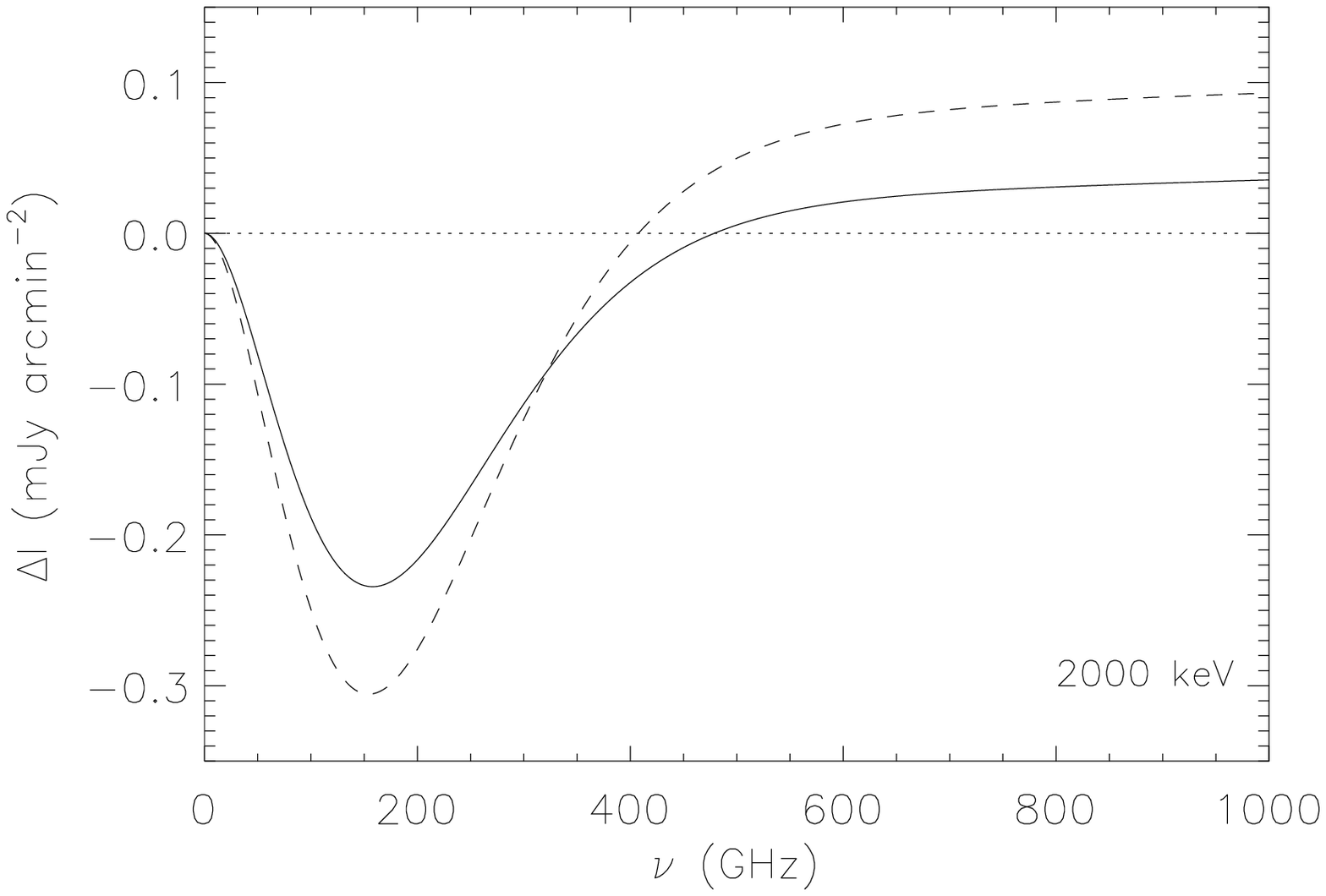} \\
\end{tabular}
\caption{Comparison between thermal (solid line) and non-thermal (dashed line) SZ spectra assuming four different values for the temperature of the thermal gas inside the bubble, as reported in the labels inside the plots; electrons parameters are reported in Table \protect\ref{tab.sz.th}. The dotted line indicates the zero level.}
\label{fig.sz.th.nt}
\end{figure*}

In Table \ref{tab.sz.th} we report the properties of the thermal and non-thermal populations for each value of the temperature of the thermal gas. As it is possible to see, higher temperatures require lower thermal densities, and as a consequence they imply higher optical depths for the non-thermal component. The consequences of this fact on the SZ spectrum are shown in Fig.\ref{fig.sz.th.nt}, where the thermal and non-thermal SZ spectra are plotted.
 
\begin{table*}{}
\caption{Values of the assumed thermal gas temperature, and corresponding values of thermal density and optical depth, non-thermal optical depth, pressure, and pseudo-temperature, and thermal and non-thermal crossover frequencies of the SZ effect.}
\begin{center}
\begin{tabular}{|*{8}{c|}}
\hline 
$(k_BT_e)_{th}$ & $n_{th}$ & $\tau_{th}$ & $\tau_{nt}$ & $P_{nt}$ & $\langle k_B T_e \rangle_{nt}$ & $\nu_{0,th}$ &  $\nu_{0,nt}$\\
keV & cm$^{-3}$&  & &  keV cm$^{-3}$ & keV & GHz & GHz \\
\hline 
500 & $7.50\times10^{-5}$ & $3.08\times10^{-5}$ & $4.58\times10^{-6}$ & $2.57\times10^{-3}$ & 2245 & 350 & 418\\
1000 & $3.75\times10^{-5}$ & $1.54\times10^{-5}$ & $7.72\times10^{-6}$ & $3.67\times10^{-2}$ & 1903 & 409 & 412\\
1500 & $2.50\times10^{-5}$ & $1.03\times10^{-5}$ & $9.63\times10^{-6}$ & $4.28\times10^{-2}$ & 1780 & 449 & 409\\
2000 & $1.88\times10^{-5}$ & $7.73\times10^{-6}$ & $1.09\times10^{-5}$ & $4.67\times10^{-2}$ & 1715 & 479 & 408\\
\hline
 \end{tabular}
 \end{center} 
 \label{tab.sz.th}
 \end{table*} 
 
It is possible to see that for temperatures until 1000 keV the thermal SZ effect dominates on the non-thermal one, both at low and high frequencies. For 1500 keV the two effects are similar, with the thermal effect dominating on the non-thermal one at low frequencies, and the non-thermal one dominating at high frequencies, and with a frequency window, between 409 and 449 GHz, where the thermal effect is negative, while the non-thermal effect is positive (see last two columns in Table \ref{tab.sz.th}, where the crossover frequencies are reported). 
For 2000 keV the non-thermal effect is more intense than the thermal one also at low frequencies, with the thermal effect being negative until 479 GHz, unlike the non-thermal one, which has a crossover frequency at 408 GHz. 
In general in the frequency range between the respective crossover frequencies the thermal and non-thermal signals have opposite signs, and this property in principle can be used to distinguish between the two cases.
 
It is also interesting to note that, as it is possible to see by comparing the values of the temperature of the thermal gas and the pseudo-temperature of non-thermal electrons, the two values are comparable for temperatures between 1500 and 2000 keV; an apparently paradoxical consequence is that, for gas temperatures higher than this value, in the non-thermal SZ spectrum the relativistic effects (as the position of the crossover frequency, which is shifted towards higher frequencies because of relativistic effects) are less strong than in the high-temperature thermal SZ spectrum. We also observe that the values of the crossover frequency are already similar for a temperature of the order of 1000 keV, when the non-thermal pseudo-temperature is still about twice the thermal temperature, indicating that the pseudo-temperature parameter of a non-thermal electrons population is not sufficient, by itself, to uniquely quantify the importance of relativistic effects in the SZ spectrum when compared to the temperature of a thermal electrons population.
 
By looking at the values of the non-thermal pressure reported in Table \ref{tab.sz.th}, it is possible to see that the non-thermal pressure is lower than the thermal one from the external ICM, which is equal to $3.75\times10^{-2}$ keV cm$^{-3}$,  
for the cases with 500 and 1000 keV, being quite similar in this last case; the non-thermal pressure is instead bigger than the thermal one for the cases with 1500 and 2000 keV, by a factor of the order of 1.25 in the last case. This would mean that in these cases the relativistic bubble would be overpressurized compared to the external ambient, and therefore that the expansion phase would not be finished yet; this would be in line with the assumption of an expansion still going on that we implicitly made in Section 2, when we have assumed an expansion exponent $q$ constant until present time.

Another interesting consideration is that for the case with 1000 keV the non-thermal and the thermal pressures inside the bubble are quite similar, but the thermal SZ effect results to be quite stronger than the non-thermal one along the whole spectral range. 
This is in apparent contradiction with the common assumption according to which the SZ effect is a direct measure of the pressure of a plasma. The explanation of this fact can be understood by writing the relativistic SZ effect as $\Delta I(x)\propto y \tilde{g}(x)$, where, in analogy with the non-relativistic formulation, $y$ contains the integral along the line of sight of the plasma pressure, and $\tilde{g}(x)$ is the function giving the frequency dependence. This function, unlike in the non-relativistic case, depends on the properties of the electrons population, like the temperature for a thermal gas or the spectral shape for a non-thermal population (e.g. Colafrancesco et al. 2003). The point is that for very high temperature populations, or for non-thermal electrons, the changes in the function $\tilde{g}(x)$ with the electrons properties can be so high that the resulting SZ effect depends in a stronger way on $\tilde{g}(x)$ than on $y$. Therefore, we can point out that when the relativistic effects are very important (non-thermal electrons or very high temperature gas) it is not true that the intensity of the SZ effect is a direct measure of the pressure of the plasma.

\section{Comparison with the SZ effect from the external ICM}

In this section we compare the non-thermal SZ effect produced inside the bubble with the thermal SZ effect produced by the external ICM. We model this signal by assuming the distribution of the ICM density and temperature as derived from X-ray measures, and exclude from the computation the volume occupied by the bubble (Pfrommer et al. 2005).

We model the density of the ICM in \MS7 using a double $\beta$-model,
\begin{equation}
n_e(r)=n_{e1}\left[1+\left(\frac{r}{r_{c1}}\right)^2\right]^{-\frac{3}{2}\beta_1}+n_{e2}\left[1+\left(\frac{r}{r_{c2}}\right)^2\right]^{-\frac{3}{2}\beta_2},
\end{equation}
and fit the function parameters to the deprojected density profile reported in fig.7 in Vantyghem et al. (2014). We obtain the values $n_{e1}=5.1\times10^{-2}$ cm$^{-3}$, $n_{e2}=6.4\times10^{-3}$ cm$^{-3}$, $r_{c1}=75$ kpc, $r_{c2}=225$ kpc, $\beta_1=3.0$, and $\beta_2=0.8$. 
 
We calculate the optical depth of the ICM along a line of sight at the distance of 150 kpc from the cluster center, where the bubbles centers are approximately located, excluding from the integration along the line of sight the region of central 200 kpc, corresponding to the diameter of the bubble. Since it has been found that to account for $\sim95\%$ of the cluster SZ signal it is necessary to integrate out to the radius of $4R_{200}$ (Battaglia et al. 2010), we integrate until this radius, adopting as an estimate of $R_{200}$ the virial radius estimated by Gitti et al. (2007), $R=2.23$ Mpc. With these assumptions, we obtain an optical depth in direction of the bubble center of $4.0\times10^{-3}$. We calculate the thermal SZ effect using this value of the optical depth and a temperature of the gas of 5.5 keV, as derived from the deprojected temperature profile at a distance of 150 kpc from the cluster center (see fig.7 in Vantyghem et al. 2014). The cool core region of the cluster extends until 100 kpc, so it should not affect the SZ effect at the distance of 150 kpc from the cluster center.

 \begin{figure*}
\centering
\begin{tabular}{cc}
\includegraphics[width=\columnwidth]{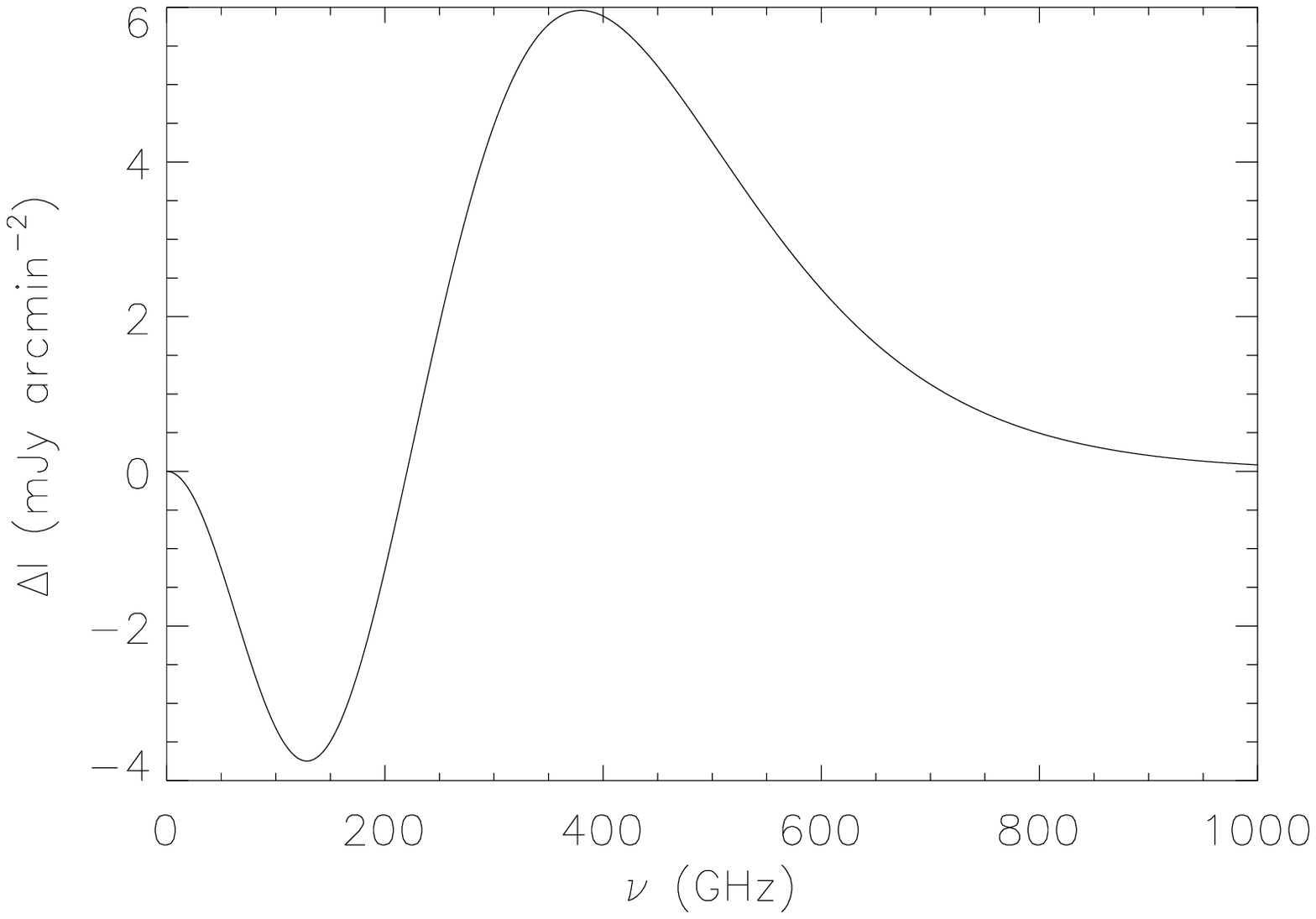} &
\includegraphics[width=\columnwidth]{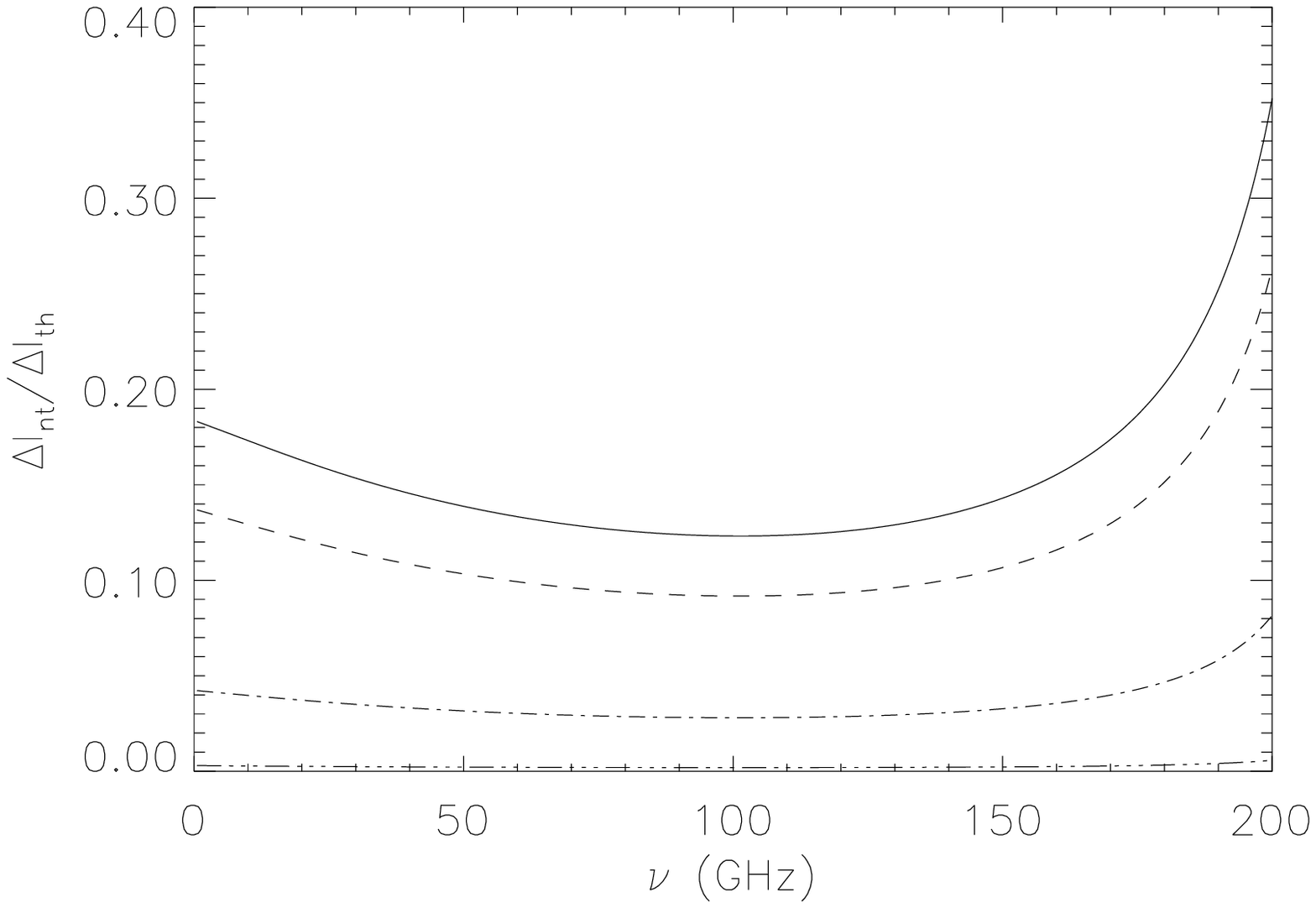} \\
\includegraphics[width=\columnwidth]{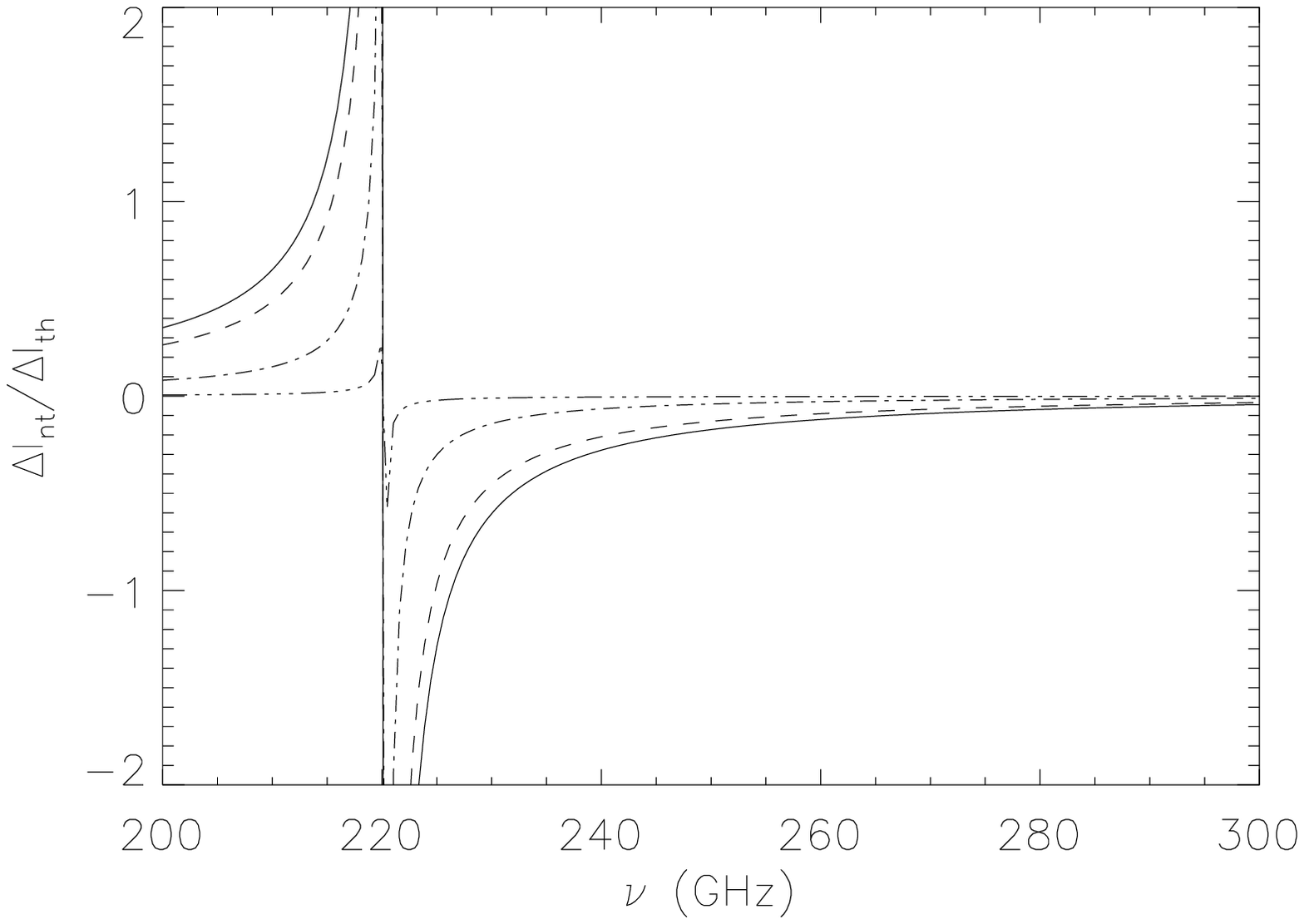} &
\includegraphics[width=\columnwidth]{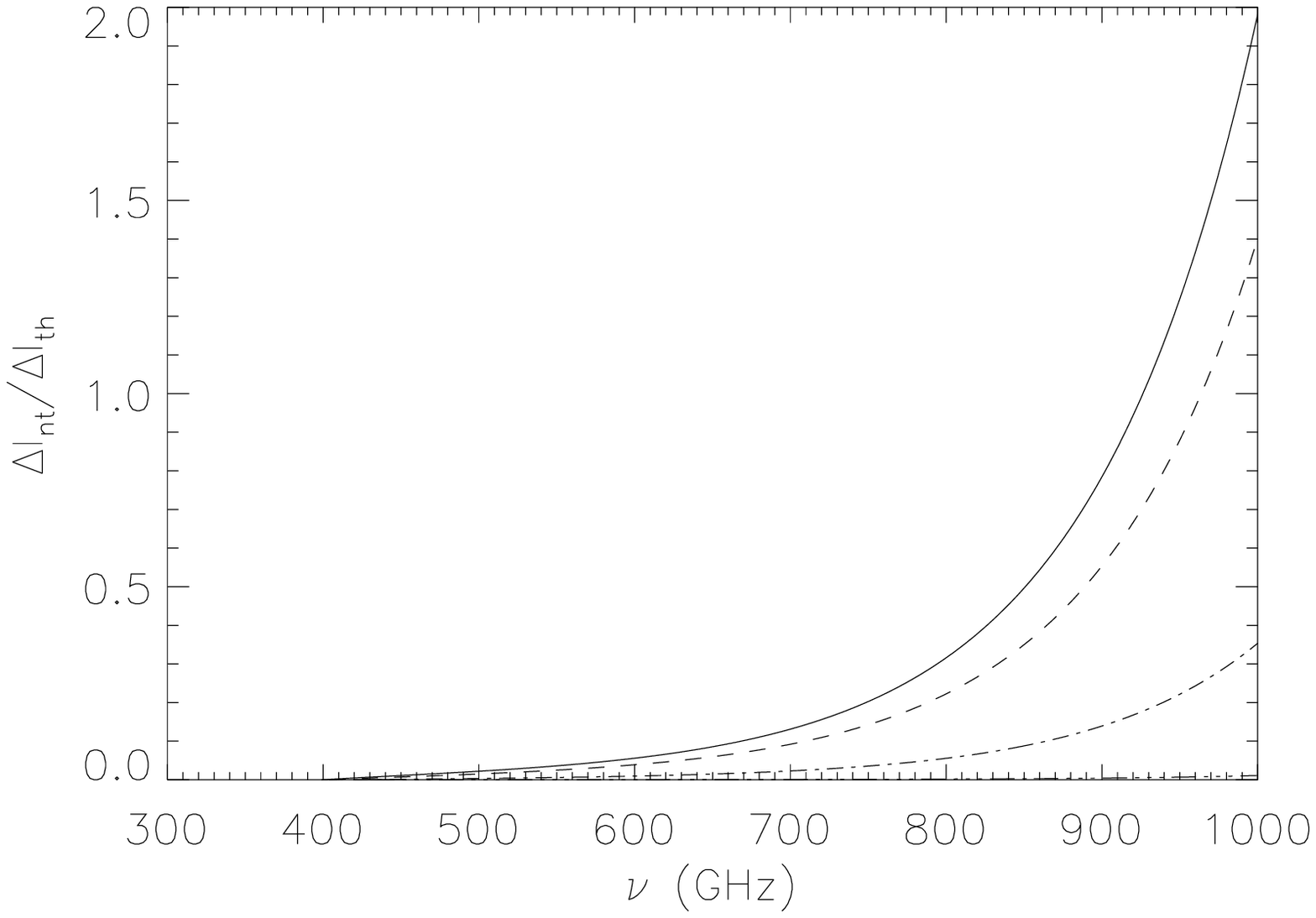} \\
\end{tabular}
\caption{First panel: thermal SZ effect produced by the external ICM in the direction of the bubble center. Other panels: ratio between the non-thermal SZ effect inside the bubble (see Fig.\protect\ref{fig.sz}, with the same meaning of line styles) and the thermal SZ of the external ICM in three different frequency bands.}
\label{fig.sz.comp}
\end{figure*}

The resulting SZ spectrum is shown in the first panel of Fig.\ref{fig.sz.comp}. This signal is stronger than the non-thermal effect expected inside the bubble (see Figs. \ref{fig.sz} and \ref{fig.sz.th.nt}), by a factor that can be of the order of 10 or higher in correspondence of the minimum and maximum frequencies (around 128 and 380 GHz respectively for a thermal gas with a temperature of 5.5 keV), indicating that the detection of the non-thermal SZ effect in the bubble is a hard task. 

In order to determine at which frequencies the possibility to detect an SZ effect of non-thermal origin is higher, in the other three panels of Fig.\ref{fig.sz.comp} we show the ratio between the non-thermal SZ effect inside the bubble and the thermal SZ effect of the external ICM in three frequency bands. 

In the low frequency band (0--200 GHz), which includes the minimum of the thermal effect, the ratio is of the order of 10--15\% for the cases where the thermal density inside the bubble is $10^{-6}-10^{-5}$ cm$^{-3}$, around 5\% for $10^{-4}$ cm$^{-3}$, and very low for $10^{-3}$ cm$^{-3}$; the ratio is minimum in correspondence of the minimum of the thermal effect. 

In the intermediate frequency range (200--300 GHz), where the crossover frequency of the thermal effect is located (at 220 GHz), the ratio has a sharp increase close to this crossover frequency, and is negative at higher frequencies; this is because for the non-thermal SZ effect the crossover frequency is located at higher frequencies (399, 405, 422, and 488 GHz for $10^{-6}$, $10^{-5}$, $10^{-4}$, and $10^{-3}$ cm$^{-3}$ respectively), and until these frequencies thermal and non-thermal effects have opposite signs. 

The frequency range where the ratio reaches higher values (excluding the very narrow interval around the thermal crossover frequency) is at high frequencies (500--1000 GHz). For a thermal density inside the bubble of $10^{-6}-10^{-5}$ cm$^{-3}$, the ratio is of the order of 10\% after 600 GHz, 30--40\% around 800 GHz, and higher than 1 after 900--950 GHz. For a value of the thermal density of $10^{-4}$ cm$^{-3}$, the ratio is about 10\% at 800 GHz and reaches 20\% around 950 GHz. For $10^{-3}$ cm$^{-3}$, the ratio is very low along the whole considered spectral range.

\section{Discussion and conclusions}

The results of this paper indicate that the intensity of the non-thermal SZ effect produced inside the bubbles of relativistic plasma in galaxy clusters is related to the density of the thermal gas inside the same bubbles, because Coulomb losses determine the shape of the electrons spectrum at low energies, from which the non-thermal SZ effect is strongly dependent. 

The detection at 30 GHz of the SZ effect in the cavities of the cluster \MS7 performed by Abdulla et al. (2019) was not sufficient, by itself, to determine if the main electrons population inside the bubble is thermal or non-thermal. By looking at fig.8 of that paper, it is possible to see that their results, in the case of a thermal gas, indicate that high temperatures are favored, with best fit values being higher than 1000 keV. In this paper we have found that, for temperatures of the order of 1500 keV, thermal and non-thermal SZ effects are expected to have similar intensities along the whole spectral range but slightly different spectral shapes, with the thermal effect being more intense in the negative part of the spectrum, and the non-thermal effect being more intense in the positive part,
indicating a possible way to discriminate between the two cases. Moreover, at high frequencies the non-thermal effect is expected to be a significant fraction of the thermal effect from the surrounding ICM, suggesting that a detection is a difficult task, but possible in principle.

Since the angular size of the cavities in \MS7 is of the order of 1 arcmin, an angular resolution smaller than this value is necessary to detect their signal without diluting it with the surrounding ICM. While at low frequencies (150 and 260 GHz) this resolution can be provided by an instrument like NIKA2 (Adam et al. 2018; Perotto et al. 2020), at higher frequencies this resolution is not provided by the Planck satellite, while the high frequency channel of Olimpo (480 GHz) has an angular resolution close to 1 arcmin (Coppolecchia et al. 2013), so it should be at the limit for this detection. The ideal instrument for this purpose is the planned space telescope Millimetron, which is expected to have an angular resolution of 6 arcsec at 1000 GHz (Kardashev et al. 2014). 

We point up that there is a number of astrophysical effects that can make it problematic a detection of the SZ effect produced in the bubbles and the determination of the nature of their content from this kind of observations.

\begin{figure}
\centering
\begin{tabular}{c}
\includegraphics[width=\columnwidth]{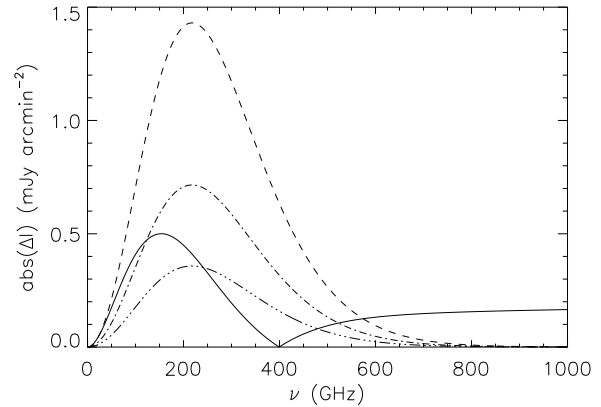}
\end{tabular}
\caption{Absolute value of the spectrum of the non-thermal SZ effect produced in the case of  thermal density  $10^{-6}$ cm$^{-3}$ (solid line), compared with the kinetic effect produced in the bubble direction by the surrounding ICM with different values of the cluster peculiar velocity: 1000 (dashed line), 500 (dot-dashed), and 250 (three dots-dashed) km s$^{-1}$.}
\label{fig.kinsz}
\end{figure}

A possible contaminant for these measures is the kinetic component of the SZ effect, produced by the peculiar motion of the cluster along the line of sight (Sunyaev \& Zel'dovich 1972). Due to its spectral shape different from the thermal one, this component can modify the overall spectrum, also shifting the crossover frequency to higher values, similarly to what relativistic effects in high temperature thermal gas or non-thermal electrons populations can do. Pfrommer et al. (2005) discussed the role of the kinetic SZ effect produced by the cluster motion in the context of the SZ effect from bubbles, showing how the kinetic effect introduces an additional unknown parameter (the peculiar velocity), which can modify the SZ effect from the whole cluster (including in the direction of the bubbles), and as a consequence can make it more difficult to estimate the contribution due to the bubble. To have an idea of the amount of the kinetic SZ contribution, we compare in Fig.\ref{fig.kinsz} the non-thermal SZ effect in the most favorable case we have considered, i.e. the one with thermal density inside the bubble of $10^{-6}$ cm$^{-3}$, with the kinetic SZ effect calculated for an optical depth corresponding to the one of the external ICM in direction of the cavity center, calculated to be $4.0\times 10^{-3}$ in Section 5, and for three different values of the peculiar velocity, 1000, 500, and 250 km s$^{-1}$. The non-thermal SZ effect is expected to be bigger than the kinetic one at high frequencies ($>600$ GHz in this case), while at lower frequencies it can have similar of lower intensity compared to the kinetic one. Since the kinetic SZ effect should be present in the whole cluster, by using instruments able to detect the SZ effect in small regions of the cluster thanks to their high spatial resolution and sensitivity like Millimetron, one can think to estimate the entity of the kinetic SZ contribution from the SZ effect measured in regions of the cluster sufficiently far from the bubbles, where the deviations from the thermal effect can be assumed as due only to the kinetic effect.

As pointed out by Ehlert et al. (2019), also the bubbles themselves can produce a kinetic SZ effect if the jets are not perpendicular to the line of sight; therefore also this effect needs to be considered as a possible contribution to the total SZ effect from the bubbles. As discussed by the same authors, this effect can be addressed using clusters where the direction of the jets appears to be close to be perpendicular to the line of sight (as probably in the case of \MS7), in order to reduce the velocity component of the jets along the line of sight, and averaging the signal of the two bubbles located in opposite directions from the cluster center, in order to cancel (or at least reduce) the combined signal if the velocities of the two jets are oriented in opposite directions. We also note that the kinetic SZ is proportional to the optical depth, which inside the bubbles is expected to be of the order of $10^{-7}-10^{-5}$ (see Table \ref{tab.sz}), i.e. smaller than the optical depth of the cluster by a factor bigger than 100 or higher; as a consequence, it can be expected that the kinetic SZ due to the bubbles should not be higher than the kinetic SZ effect of the whole cluster.

By looking at Figs.\ref{fig.sz.comp} and \ref{fig.kinsz}, the immediate deduction would be that the optimal frequency range where to observe the SZ effect from the bubbles would be the high-frequency range ($\nu>600$ GHz), where the ratio between the SZ effect inside the bubble and the external one is maximized, and where the bias produced by a possible kinetic component is reduced.
However, at these frequencies observations are strongly challenged by infrared emission from different kinds of astrophysical sources. The foreground emission from Galactic interstellar dust (e.g. Planck Collaboration 2014) is a main issue in the high frequencies range, because it can overwhelm the intensity of the SZ signal after $\sim500$ GHz (e.g. de Bernardis et al. 2012), making it necessary to use specific techniques to model the dust emission and extract the SZ signal (e.g. Meisner \& Finkbeiner 2015; Bourdin et al. 2017). While the spectral shape of the Galactic dust signal can be successfully modeled with one or two black body functions multiplied by power-law functions (Finkbeiner, Davis \& Schlegel 1999), the spatial fluctuations of the dust density and temperature make it very hard to estimate the exact amount of the foreground emission at the location of the cluster. In principle, it is possible to measure the SZ effect using spectrometers in differential mode, measuring the difference of the signal between two different directions, on- and off-cluster, like it is possible for the planned LACS camera on board of Millimetron; in this way at least the foreground emission fluctuating on large scale should be eliminated (see e.g. discussions in Cooray 2006 and Colafrancesco, Marchegiani \& Emritte 2016); however, fluctuations on smaller scale can still be present making it difficult to estimate the SZ signal at high frequencies. Other possible sources of infrared emission are the dust emission inside the cluster itself, which has been found to be present in galaxy clusters from the analysis of stacked samples (Montier \& Giard 2005), and background galaxies at high redshifts; to address these problems, it can be useful to study the cluster at higher frequencies with infrared instruments, in order to estimate their amount and model their spatial and spectral shapes at the frequencies of interest. Also in this case it would be important to detect the SZ emission with instruments with high angular resolution, because if the infrared emission is detected only in the direction of the bubbles, it can be produced by the SZ effect in the bubbles, which is expected to continue after 1000 GHz, while if the infrared emission is detected as a diffuse emission in the whole cluster, it would be more reasonable to conclude that it is due to the dust content of the cluster. 

We can note that the results of this paper have been obtained with a number of assumptions and approximations.
On one hand, the methods followed in this paper for describing the evolution of the electrons spectrum in the bubbles are an advancement compared to previous papers 
(e.g. Kaiser, Dennett-Thorpe \& Alexander 1997; Acharya, Majumdar \& Nath 2020), where it was used an analytical solution including the effect of radiative losses and adiabatic expansion, but not Coulomb losses and Fermi-II acceleration, which are instead included in this paper, by performing a numerical computation. 
On the other hand, in this calculation we have made several assumptions, as
the one of spherical symmetry for the cavities.  
Regarding this point, we can observe that both X-ray (Vantyghem et al. 2014) and SZ analysis (Abdulla et al. 2019) indicate that the shape of the bubbles in \MS7 is elliptical. Using the values in table 2 in Vantyghem et al. (2014) for the outer south-western bubble (the most elliptical one), and assuming that the size of the bubble along the line of sight is equal to the one on the minor axis (for that bubble the minor semiaxis is equal to 100 kpc, i.e. the same value we have assumed as the bubble radius in the spherical approximation), it can be estimated that the volume of the bubble calculated using the elliptical assumption is 1.2 times the volume calculated with the spherical assumption. As a consequence, the density of non-thermal electrons, which has been normalized to the observed radio emission, would be lower by the same factor, and therefore the non-thermal SZ effect calculated with the spherical assumption might be overestimated by the same factor. However, error bars in table 2 in Vantyghem et al. (2014) are quite large, so it is difficult to have an exact idea of the amount of this factor. 

Another assumption is the one of constant thermal and non-thermal densities, and constant magnetic field intensity inside the bubbles. In this respect, we note that available radio data are given as fluxes integrated over the bubble region, and therefore do not allow to estimate the spatial distribution of electrons and magnetic field. We also note that in this paper we have calculated the SZ effect in the direction of the bubble center, without taking into account the finite instrument angular resolution. Specific predictions for real instruments will need to take into consideration the instrument resolution, and the spatial distribution of the SZ signal inside the bubble.

Moreover we have also modeled the ICM as a spherical unperturbed system with a hole in correspondence of the cavity, whereas it has been found that the expansion of the lobe inside the ICM gives origin to a shock front, producing an increase of the temperature and the density of the ICM close to the external edge of the cavities, which can enhance the external SZ effect also along the line of sight (Ehlert et al. 2019). For this reason, it is possible that the thermal SZ effect from the ICM calculated in this paper is underestimated, and that therefore the estimates presented in Fig.\ref{fig.sz.comp} are too optimistic.

For all these reasons, the conclusions of this paper can be considered as a proof of concept of the relationship between the density of the high-temperature thermal gas inside X-ray cavities in galaxy clusters and the shape of the low-energy spectrum of the electrons located in the same cavities, studying the consequences on the properties of the non-thermal SZ effect. We have shown that, since when the gas temperature is high its density is expected to be low, for high gas temperatures the non-thermal SZ effect is expected to be stronger compared to the thermal SZ effect in the cavity. We have also shown that the non-thermal SZ effect in most favorable cases (with density of the high-temperature gas of the order of $10^{-5}$ cm$^{-3}$ or less) can be a non-negligible fraction of the thermal effect produced by the external ICM in the cavity direction, especially in the high-frequencies band. However, for more accurate quantitative results, more realistic models  
for describing the properties of the cavities and the cluster, and for addressing observational contaminations, will be necessary.


\section*{Acknowledgments}
The author thanks the reviewer for useful comments and suggestions.

\section*{Data availability}
No new data were generated or analysed in support of this research.



\bsp

\label{lastpage}

\end{document}